\documentclass[
reprint,
 superscriptaddress,
 amsmath,amssymb,
 aps,
prb,
]{revtex4-2}
\usepackage{tabularx}
\usepackage[colorlinks,bookmarks=false,citecolor=darkblue,linkcolor=red,urlcolor=blue]{hyperref}
\usepackage{graphicx}
\usepackage{dcolumn}
\usepackage{bm}
\usepackage{multirow}
\usepackage{booktabs}
\usepackage[dvipsnames]{xcolor}
\usepackage{upgreek}
\usepackage{boldline}
\usepackage{bigdelim}
\usepackage{xcolor}

\definecolor{darkblue}{rgb}{0,0.02,0.45}

\newcommand{\fmo}{\ce{Fe2Mo3O8}}

\newcommand{\mmo}{\ce{Mn2Mo3O8}}
\newcommand{\cmo}{\ce{Co2Mo3O8}}
\newcommand{\TN}{\ensuremath{T_{\mathrm{N}}}}

\newcommand{\Epara}{$\mathbf{E}^{\omega}\parallel c$}

\usepackage[warn]{textcomp}
\usepackage{mathtools}

\DeclarePairedDelimiterX\braket[2]{\langle}{\rangle}{#1 \delimsize\vert #2}

\usepackage[version=4]{mhchem}
\usepackage{mathptmx}

\usepackage{amsmath}

\begin{document}

\title{Magnetic circular dichroism of THz modes and selection rules of Raman-active optical phonons in the polar altermagnet candidate \mmo}

\author{F.~Schilberth}
\affiliation{Experimental Physics V, Center for Electronic
Correlations and Magnetism, Institute for Physics, University of Augsburg, D-86159 Augsburg, Germany}
\author{K.~Vasin}
\affiliation{Experimental Physics V, Center for Electronic
Correlations and Magnetism, Institute for Physics, University of Augsburg, D-86159 Augsburg, Germany}
\affiliation{Institute for Physics, Kazan (Volga region) Federal University, 420008 Kazan, Russia}
\author{M.~Knauft}
\affiliation{Max Planck Institute for Solid State Research, Heisenbergstrasse 1, 70569 Stuttgart, Germany}
\author{M. Kond\'{a}kor}
\affiliation{Department of Theoretical Physics, Institute of Physics, Budapest University of Technology and Economics, M\H{u}egyetem rkp. 3., H-1111 Budapest, Hungary}
\affiliation{Institute for Solid State Physics and Optics, HUN-REN Wigner Research Centre for Physics, H-1525 Budapest, P.O.B. 49, Hungary}
\author{M.~Vuckovic}
\author{N.~Herrmann}
\affiliation{Experimental Physics V, Center for Electronic
Correlations and Magnetism, Institute for Physics, University of Augsburg, D-86159 Augsburg, Germany}

\author{K. Penc}
\affiliation{Institute for Solid State Physics and Optics, HUN-REN Wigner Research Centre for Physics, H-1525 Budapest, P.O.B. 49, Hungary}
\author{M.~Minola}
\author{B.~Keimer}
\affiliation{Max-Planck-Institute for Solid State Research, Heisenbergstrasse 1, 70569 Stuttgart, Germany}
\author{V.A.~Martinez}
\author{K.~Panda}
\author{A.A.~Sirenko}
\affiliation{Department of Physics, New Jersey Institute of Technology, Newark, New Jersey 07102, USA}
\author{L.~Prodan}
\affiliation{Experimental Physics V, Center for Electronic
Correlations and Magnetism, Institute for Physics, University of Augsburg, D-86159 Augsburg, Germany}
\author{V.~Tsurkan}
\affiliation{Experimental Physics V, Center for Electronic
Correlations and Magnetism, Institute for Physics, University of Augsburg, D-86159 Augsburg, Germany}
\affiliation{Institute of Applied Physics, Moldova State University, str. Academiei 5, MD2028, Chisinau, Republic of Moldova}

\author{A.A.~Tsirlin}
\affiliation{Felix Bloch Institute for Solid-State Physics, Leipzig University, 04103 Leipzig, Germany}
\author{I.~K\'ezsm\'arki}
\affiliation{Experimental Physics V, Center for Electronic
Correlations and Magnetism, Institute for Physics, University of Augsburg, D-86159 Augsburg, Germany}
\author{J.~Deisenhofer}
\affiliation{Experimental Physics V, Center for Electronic
Correlations and Magnetism, Institute for Physics, University of Augsburg, D-86159 Augsburg, Germany}

\date{\today}

\begin{abstract}

We investigated the magnetic and vibrational excitations in the collinear altermagnet candidate \mmo{} by temperature dependent Raman scattering and magneto-optical THz time-domain transmission spectroscopy. By comparison to \textit{ab initio} calculations accurately capturing the eigenfrequencies of the vibrational eigenmodes, we identify all optical phonons, including the lowest-lying Raman modes of $A_1$ and $E_2$ type, which had remained elusive in a previous Raman study. Moreover, we compare the selection rules for optically active phonons in the paramagnetic and the magnetically ordered phases of \mmo{} and analyze the Raman selection rules with respect to pseudo-angular momentum conservation. No evidence of the expected splitting of the degenerate paramagnetic $E_2$ optical phonons into  modes with circular polarization upon magnetic ordering could be resolved, likely due to weak spin-orbit coupling typical for Mn$^{2+}$. In contrast, we observe strong magnetic circular dichroism at a broad THz excitation band, emerging in the magnetically ordered state. This band, potentially originating from two-magnon excitations, is only electric-dipole active and features a field-dependent two-component fine structure. Its magnetic circular dichroism vanishes above the spin-flop transition at 4~T.

\end{abstract}

\maketitle

\section{Introduction}
\begin{figure}[t]
    \centering
    \includegraphics[width=\linewidth]{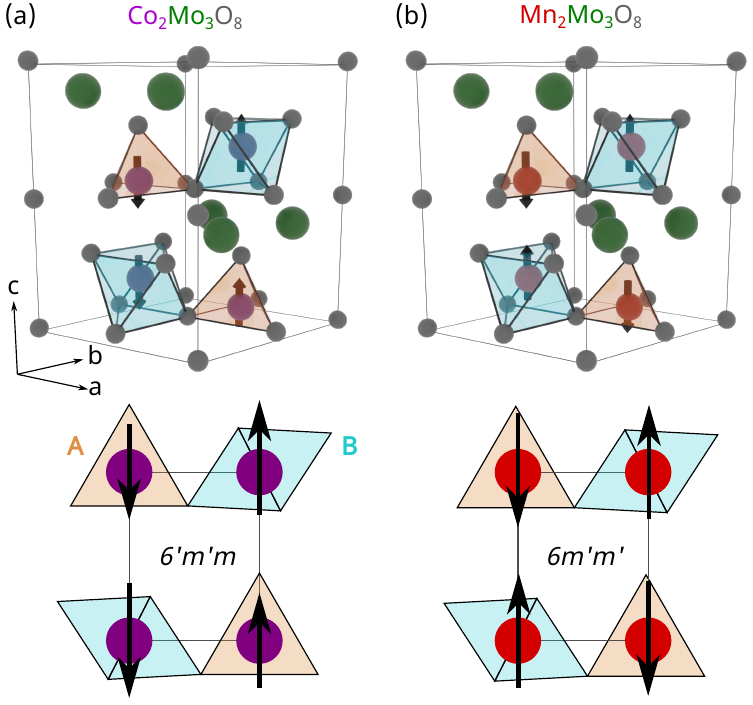}
    \caption{Comparison of the different magnetic structures of (a) \cmo/\fmo{} and (b) \mmo. While for \cmo/\fmo{} the compensated antiferromagnetic state comprises four magnetic sublattices within the magnetic point group $6^\prime m^\prime m$, \mmo{} is described by two magnetic sublattices within the magnetic point group $6 m^\prime m^\prime$, which allows for ferrimagnetism.}
    \label{fig:Symmetries}
\end{figure}

The polar molybdenum oxides \ce{A2Mo3O8}  (A = Mn, Fe, Co) exhibit collinear magnetic ground states, which can be changed by external magnetic fields or doping \cite{Kurumaji:2015,Wang:2015,Kurumaji:2017,Tang:2019,Csizi:2020,Tang:2022,Reschke:2022,Prodan:2022,Ghara:2023,Szaller:2025}. In the paramagnetic regime, these materials exhibit the non-symmorphic polar hexagonal space group $P6_3mc$ (\#186) with an intrinisic polarization along the $c$-axis. The structural unit cell is shown in Fig.~\ref{fig:Symmetries}. The A$^{2+}$ ions are responsible for magnetism, with half of them occupying the corner-sharing tetrahedral (A) and the other half the octahedral (B) sites. The Mo ions build non-magnetic trimers \cite{Varret:1972,Cotton:1964,Wang:2015}.

\begin{figure}[t]
    \centering
    \includegraphics[width=\linewidth]{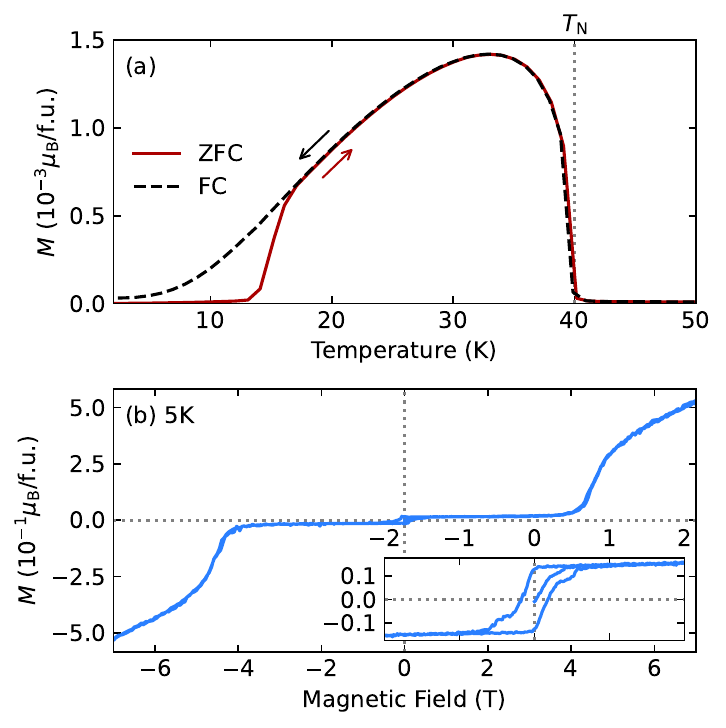}
    \caption{(a) Magnetization of \mmo{} as a function of temperature measured upon heating after zero-field cooling (ZFC) and upon cooling in field (FC), both in a field of $H=100$~Oe along the $c$-axis. (b) Magnetization as a function of magnetic field along the $c$-axis. The inset zooms in on the   hysteresis at low fields.}
    \label{fig:Magnetisation}
\end{figure}

As shown in Fig.~\ref{fig:Symmetries}(a), \cmo{} and \fmo{} exhibit collinear AFM order along the $c$-axis below $\TN{}=39$~K \cite{Tang:2019,Reschke:2022,Tang:2022,Prodan:2022, Szaller:2025} and $\TN{}=60$~K \cite{Kurumaji:2015,Wang:2015}, respectively. The magnetic structure of \mmo{} below $\TN{}=40$~K corresponds to a ferrimagnetic-like configuration with two ferromagnetic sublattices, one residing on the tetrahedral A-site, the other on the octahedral B-sites (see Fig.~\ref{fig:Symmetries}(b)) \cite{Szaller:2020,Liao:2025}.
The temperature dependence of the magnetization in \mmo{} is shown in Fig.~\ref{fig:Magnetisation}(a). The magnetization increases below $T_\text{N}=40$~K, which had been attributed to a slightly canted spin state and different temperature dependences of the sublattice magnetization on A and B sites \cite{Szaller:2020,Liao:2025}. Towards lowest temperatures, these two sublattices completely compensate each other leading to a state with magnetization $\mathbf{M}\rightarrow 0$ for $T\rightarrow 0$ \cite{McAlister:1983}. The magnetization as a function of magnetic field, shown in Fig.~\ref{fig:Magnetisation}(b) at 5~K, identifies the onset of the spin-flop transition at about 4~T and exhibits a small hysteresis in low fields, which indicates the formation of ferrimagnetic domains in the magnetically ordered state (see inset of Fig.~\ref{fig:Magnetisation}(b)). These domains are aligned in fields exceeding 1~T at 5~K in agreement with previous reports \cite{Szaller:2020}.

The magnetic symmetries of these materials allow the observation of very interesting magneto-optical and magnetoelectric effects: For \fmo{}, low-lying chiral phonons and magnon-polariton excitations in the THz frequency range have been reported \cite{Kurumaji:2017a,Wu:2023,Bao:2023,Vasin:2024}, which exhibit non-reciprocal directional dichroism \cite{Reschke:2022,Vasin:2024}. \mmo{} was shown to exhibit a linear magnetoelectric effect \cite{Kurumaji:2017}. 

Recently, it was recognized that the three compensated collinear magnets \fmo, \cmo, and \mmo{} fulfill the symmetry criteria for altermagnets \cite{Cheong2024,Chen:2025}, which can exhibit spin splittings of electronic bands or splittings of magnons along general directions in the Brillouin zone in a non-relativistic setting \cite{Smejkal:2022,McClarty:2025}. In a detailed study of optical phonons in the paramagnetic and magnetically ordered state of \cmo{}, the selection rules imposed by magnetic point groups and the proposed spin group approach for altermagnets were compared for infrared (IR)- and Raman-active phonons \cite{Schilberth:2026}. As a result, \cmo{} proved to be a prominent realization of changes in the phonon sector due to magnetic order as described by Anastassakis and Burstein \cite{Anastassakis:1972} using magnetic point groups.
More generally, the change of phonon selection rules due to magnetic ordering or the application of external fields such as electric, magnetic, or strain fields have been described as \textit{morphic effects} by Anastassasakis and Burstein in a series of pioneering papers in the 1970s \cite{Anastassakis:1971,Anastassakis:1971a,Anastassakis:1972a,Anastassakis:1972b,Anastassakis:1972}, where the term morphic is related to the symmetry-adapted shape of the effective-charge tensor of the material in the presence of external fields or magnetic order.
In contrast, the spin group approach for ideal altermagnets with $\mathbf{q}=0$ antiferromagnetic order predicts, that there should be no changes with respect to the selection rules of optical phonons at the $\Gamma$ point upon magnetic ordering  \cite{Schilberth:2026}.


Here, we follow a similar route for \mmo{}, where the magnetic point group symmetry of the collinear magnetic state imposes a splitting of the doubly-degenerate $E$-type phonons into conjugate left- and right-handed circularly polarized modes. Additionally, the previously silent modes become Raman- and infrared (IR) active below $T_\text{N}$. In order to reveal possible morphic effects, we investigated \mmo{} single crystals using THz-time domain and Raman spectroscopy and compared the observed modes to the eigenfrequencies obtained via \textit{ab initio} calculations. The optical phonons of \mmo{} had been investigated by IR-reflectivity and Raman scattering across the magnetic ordering transition previously, without taking into account symmetry changes governed by magnetic ordering \cite{Stanislavchuk:2020}. We additionally performed circularly polarized Raman measurements in the magnetically ordered state of \mmo{} and investigated the low-frequency IR-active excitations using THz-time domain spectroscopy in magnetic fields and circularly polarized THz radiation. 

We find that the optical Raman modes do not change upon entering into the magnetically ordered regime, i.e.\,we could not detect any changes in the phonon sector imposed by the symmetry lowering due to magnetic order. A broad THz absorption is found to consist of two components that exhibit magnetic circular dichroism. This effect disappears upon entering the spin-flop phase in fields above 4~T.

\section{Experimental and Computational Details}

Transmission measurements were performed on a thin $ab$-plane single crystal.  By using a Bruker Fourier-transform IR-spectrometer Vertex80 equipped with a He-flow cryostat, the mid-infrared (MIR) and near-infrared (NIR) frequency range from 1000 to 13000\,cm$^{-1}$ and a temperature range from 5 to 300\,K could be covered. THz time-domain spectroscopy measurements with linearly polarized light were performed in the temperature range 5-300~K  on the same crystal using a TOPTICA TeraFlash pro THz system and a He-flow cryostat in transmission geometry at the University of Augsburg.

Circularly polarized magnetic-field-dependent time-domain THz transmission spectroscopy was performed at the National Synchrotron Light Source II at Brookhaven National Laboratory using a Toptica TeraFlash Pro spectrometer in magnetic fields ranging from -7 to 7~T. The magnetic field was applied in the Faraday geometry, perpendicular to the $ab$-plane of the sample and parallel to the propagation direction of the THz beam $(\mathbf{k}\parallel\mathbf{B})$. Measurements were carried out at 5~K using a He-flow cryostat. Circular polarization was generated from broadband wire-grid linear polarizers in combination with achromatic optical retarders constructed from undoped Si prisms. These retarders transform linearly polarized light into circularly polarized states $\mathbf{e}_+=1/\sqrt{2}(\mathbf{e}_x+i\mathbf{e}_y)$ and $\mathbf{e}_-=1/\sqrt{2}(\mathbf{e}_x-i\mathbf{e}_y)$ with spin angular momenta $S_z= -\hbar$ and $S_z= +\hbar$ pointing along their propagation direction $\mathbf{k}$, yielding corresponding helicities $\sigma_\pm=\mathbf{k}\cdot\mathbf{S}/|\mathbf{k}|=\mp \hbar$, respectively. In order to define the nomenclature of left-handed and right-handed circularly polarized light, we chose the view of an observer looking in the direction of light propagation and call $\mathbf{e}_+$ as right-circularly polarized and $\mathbf{e}_-$ left-circularly polarized light \cite{Messiah:1961,Born:1999}. More experimental details on the retarders and their function can be found in Ref. \cite{Stanislavchuk:2013}.

Confocal Raman scattering experiments were performed using a LabRam HR800 spectrometer with an excitation wavelength of 633 nm from a HeNe-laser in backscattering geometry. For spectral analysis, a grating with 1800 grooves per mm was used. Both linearly or circularly polarized light was used to distinguish modes of different irreducible representations. The sample was fixed on a holder using silver paint and placed inside a He-flow cryostat. The indicated temperatures were measured at the holder position. During the experiment, the laser power was kept below 0.45 mW.

Density-functional-theory (DFT) band-structure calculations were performed in the \texttt{VASP} code~\cite{vasp1,vasp2} using the Perdew-Burke-Ernzerhof version of the exchange-correlation potential~\cite{pbe96} and the experimental lattice parameters at 1.5 K: $a = 5.7943$ \AA, $c = 10.2638$ \AA. Phonon frequencies at the $\Gamma$-point were obtained by the finite-displacement method. Electronic correlations in the Mn $3d$ shell were taken into account on the mean-field level using the DFT+$U$ procedure with the on-site Coulomb repulsion parameter $U_d=5$\,eV and Hund's coupling $J_d=1$\,eV~\cite{Szaller:2025}. The reported collinear low-temperature compensated ferrimagnetic configuration \cite{Liao:2025} was used.

\section{Symmetry analysis and selection rules for optical phonons}
\label{sec:phonons}

\subsection{Irreducible representations and corepresentations}

Phonon selection rules in nonmagnetic crystals are generally determined by the irreducible representations $\Gamma$ of the crystallographic point group.  In magnetic crystals, time reversal must also be taken into account. Unlike unitary spatial operations, time reversal is antiunitary, requiring the use of corepresentation theory. In the paramagnetic phase, the absence of static magnetic order makes the equilibrium state invariant under time reversal, and the relevant symmetry is therefore a grey magnetic group. For lattice vibrations and their conventional infrared and Raman selection rules, however, the resulting classification is equivalent to that of the underlying crystallographic point group. Once long-range magnetic order develops, the ordered moments define preferred directions in spin space. When spin and lattice degrees of freedom are coupled, symmetry operations must preserve the magnetic structure either directly or in combination with time reversal. This leads to magnetic space and point groups \cite{Bradley:2009}. Within magnetic group theory, their irreducible corepresentations, denoted by $D\Gamma$, determine the optical-phonon selection rules in the magnetically ordered phase \cite{Cracknell:1969,Anastassakis:1972,Bradley:2009,Schilberth:2026}.

For paramagnetic \mmo{}, whose crystallographic point group is $\mathbf{G}=6mm$ and whose primitive unit cell contains two formula units, the 78 normal modes at the zone center decompose into irreducible representations as
\begin{align}\label{Eq:irreps}
\Gamma &= 9A_1(z;x^2+y^2,z^2)   + 12E_1(x,y;xz,yz) &&\text{(IR + Raman)}\nonumber \\ 
&+  13E_2 (x^2-y^2,xy) &&\text{(Raman)}\nonumber \\  
&+  A_1 +E_1 &&\text{(acoustic)}\nonumber \\ 
&+ 3A_2 + 10B_1 +3B_2  &&\text{(silent),}     
\end{align}
where the corresponding symmetry-adapted basis functions are given in parentheses. 

The nine optical $A_1$ modes can therefore be identified by direct IR transmission or reflectivity measurements with \Epara{}, or by Raman scattering in the $x(zz)\bar{x}$ and $y(zz)\bar{y}$ configurations, using Porto notation \cite{Damen:1966,Rousseau:1981}, which is read from left to right starting with the propagation direction of the incoming light, then - in parentheses - the polarization state of the incoming light followed by the polarization state of the scattered light, and finally the propagation direction of the detected scattered light. Similarly, the twelve optical $E_1$ modes can be observed by IR transmission or reflectivity with $\mathbf{E}^{\omega}\parallel a$, i.e., with the electric field polarized in the hexagonal basal plane, or by Raman scattering in the $y(xz)\bar{y}$ and $x(yz)\bar{x}$ configurations. The remaining thirteen $E_2$ modes are Raman active and can be observed in the $z(xy)\bar{z}$ configuration. These polarization configurations have previously been investigated for \mmo{} and the isostructural compounds \cmo{} and \fmo{} using linearly polarized light, and the paramagnetic state selection rules were confirmed for many of the modes \cite{Reschke:2020,Stanislavchuk:2020,Schilberth:2026}. Raman measurements in the $z(xx)\bar{z}$ configuration provide an additional check because both $A_1$ and $E_2$ modes are symmetry-allowed in this channel. The intensities of the $A_1$ modes, however, need not be the same as in the $x(zz)\bar{x}$ configuration because the two geometries probe different Raman tensor elements.

Below $\TN=40$~K, long-range magnetic order develops together with a spontaneous magnetization, as shown in Fig.~\ref{fig:Magnetisation}(a). Magnetic moments canted away from the $c$-axis have been reported in the intermediate-temperature ordered phase of \mmo{} \cite{Liao:2025}, whereas a collinear magnetic structure with vanishing net magnetization is established at the lowest temperatures \cite{McAlister:1983}.
This low-temperature collinear state is described by the magnetic space group $P6_3m^\prime c^\prime$ and the corresponding magnetic point group $6m^\prime m^\prime$ \cite{Bertrand:1975,Liao:2025}.
As in \cmo{} and \fmo{}, the magnetic ordering does not enlarge the
primitive unit cell. Consequently, the number of normal modes remains 78.

To determine the phonon corepresentations of the magnetic point group, we follow the procedure of Anastassakis and Burstein \cite{Anastassakis:1972}. For \mmo{} with magnetic point group $\mathbf{M}=6m^\prime m^\prime$, the zone-center vibrational corepresentation decomposes into irreducible corepresentations as (see Tables~\ref{tab:6mm} and \ref{tab:Coreps_mmo})
\begin{align}\label{Eq:coreps}
D\Gamma &= 12DA(z;x^2+y^2,z^2) &&\text{(IR+Raman)} \nonumber\\
&+ 12(D^1E_1 + D^2E_1) (x,y;xz,yz)&&\text{(IR+Raman)}\nonumber \\ 
&+ 13(D^1E_2 + D^2E_2) (x^2-y^2;xy)&&\text{(Raman)}\nonumber \\ 
&+  DA_1 +D^1E_1 + D^2E_1&&\text{(acoustic)}\nonumber \\ 
&+ 13DB &&\text{(silent).}     
\end{align}
The corresponding symmetry-adapted linear and quadratic  basis functions are again given in parentheses. Here we follow the convention that $E_1$ representations are odd and $E_2$ are even under the 2-fold rotation $C_2$ (see Tab.~\ref{tab:6mm}). 

The magnetic group analysis predicts several morphic effects in \mmo{}. In particular, the formerly silent $A_2$ modes may become optically active in the same polarization configurations as the $A_1$ modes. This occurs because both $A_1$ and $A_2$ reduce to the same irreducible representation $A$ of the unitary halving subgroup $\mathbf{H}=6$ and therefore belong to the same corepresentation $DA$ of $\mathbf{M}=6m^\prime m^\prime$.

Furthermore, the two-dimensional irreducible representations of $6mm$ decompose into one-dimensional complex-conjugate representations of $\mathbf{H}=6$ according to
\begin{equation}
E_1 \rightarrow {}^1E_1+{}^2E_1
\quad\mathrm{and}\quad
E_2 \rightarrow {}^1E_2+{}^2E_2.
\end{equation}
They form the corresponding corepresentations in $\mathbf{M}=6 m^\prime m^\prime$. As can be seen from Tab.~\ref{tab:6mm}, the $\phantom{,}^1E_1$ $(^1E_2)$ and $\phantom{,}^2E_1$ $(^2E_2)$ representations are complex conjugates and describe circularly polarized modes (this also holds for the corepresentations). Hence, each $E_i$ $(i=1,2)$ doublet of the paramagnetic state splits into left ($D^1E_{i}$) and  right ($D^2E_{i}$) circularly polarized modes. A magnetic field applied along the sixfold axis does not further reduce the magnetic point group symmetry.

\subsection{The spin group of \mmo{} }

We next discuss the spin group symmetry of \mmo{} and assess whether its low-temperature magnetic state exhibits altermagnetic characteristics. The tetrahedral and octahedral Mn sublattices are crystallographically inequivalent and host oppositely oriented magnetic moments that become nearly compensated at low temperatures. Moreover, the magnetic coupling parameters associated with the two sublattices in an effective description have been reported to be nearly identical \cite{Liao:2025}. Because no symmetry operation connects the two magnetic sublattices, their compensation is not symmetry enforced. The low-temperature state may therefore be described as a nearly compensated ferrimagnet rather than, on this basis alone, as a conventional altermagnet. 
At the same time, the absence of a symmetry enforcing the degenerate magnon or electronic states throughout the Brillouin zone allows altermagnet-like band structures.
It is therefore interesting to study whether \mmo{} satisfies the stricter symmetry criteria for altermagnetism determined by its spin group symmetry, which we analyze below.

To identify the spin symmetry underlying the collinear magnetic state, we adopt the spin group classification of Ref.~\cite{Schiff:2025}. In the nonrelativistic limit, the spin point group of a collinear magnet can be constructed as the direct product of its collinear spin-only group $\mathbf{b}^{\infty}$ and the corresponding magnetic point group. For \mmo{}, this construction gives
\begin{equation}
    6m^\prime m^\prime \times \mathbf{b}^{\infty}
    =
    {}^{1}6^{\bar{1}}m^{\bar{1}}m,
\end{equation}
which has Litvin index 399. This spin point group differs from that of \cmo{} and \fmo{}, ${}^{\bar{1}}6^{\bar{1}}m^{1}m$ (Litvin index 396). The difference originates from their distinct magnetic point groups, $6m^\prime m^\prime$ and $6^\prime m^\prime m$, respectively: the spin-flipping operations are combined with different real-space crystallographic operations in order to preserve the corresponding magnetic structures. As shown in Ref.~\cite{Schilberth:2026}, a corepresentation analysis based on the spin group recovers, in the nonrelativistic limit, the optical-phonon selection rules of the crystallographic point group $6mm$.

\subsection{Conservation of pseudoangular momentum - Raman selection rules for phonons}

\begin{table*}[ht]
\squeezetable
\begin{ruledtabular}
\centering\footnotesize
\caption{ \label{tab:selectionrules}Selection rules for first-order Raman scattering of phonons arising from PAM conservation for the point groups $\mathbf{G}=6mm$, $\mathbf{H}=6$ and $\mathbf{M}=6m'm'$ in \mmo. The  notation $\checkmark$  and $\times$  indicates, whether a mode is allowed or forbidden by PAM and in the corresponding Raman configuration, where the PAM selection rules (columns 7-9) have to be considered in addition to the symmetry adapted functions given in Eqs.~\ref{Eq:irreps} and \ref{Eq:coreps}. The selection rules including PAM for first-order Raman processes derived using Eq.~\ref{eq:crosssec} are given in  columns 10-15 for the different Raman channels denoted in Porto notation.}
\begin{tabular}{ccc|c|cc|ccc|cccccc}
\hline
\multicolumn{2}{c}{$\mathbf{G}=6mm$} & dim & 
func. & $N_\nu$ & $m_\nu^{\mathrm{ph}}$ & \multicolumn{3}{c|}{Helicity change $\Delta \sigma$} & \multicolumn{6}{c}{Raman configuration $z(\mathbf{e_{in}}\mathbf{e_{sc}})\bar{z}$} \\
 &  & & & & & -2 & 0 & 2 & $(xx)$ & $(xy)$ & $(\sigma_{+}\sigma_{+})$ & $(\sigma_{-}\sigma_{-})$ & $(\sigma_{+}\sigma_{-})$ & $(\sigma_{-}\sigma_{+})$ \\ 
 \hline
 \multicolumn{2}{c}{$A_{1}$}  & 1 & $z;x^2+y^2,z^2$ & 6 & 0 & $\times$ & $\checkmark$ & $\times$ & $\checkmark$ & $\times$ & $\checkmark$ & $\checkmark$ & $\times$ & $\times$ \\
  \multicolumn{2}{c}{$A_{2}$}  & 1 & $-$ & 6 & 0 & $\times$ & $\checkmark$ & $\times$ & $\times$ & $\times$ & $\times$ & $\times$ & $\times$ & $\times$ \\
  \multicolumn{2}{c}{$B_{1}$}  & 1 & $-$ & 3 & 0 & $\times$ & $\checkmark$ & $\times$ & $\times$ & $\times$ & $\times$ & $\times$ & $\times$ & $\times$ \\
  \multicolumn{2}{c}{$B_{2}$}  & 1 & $-$ & 3 & 0 & $\times$ & $\checkmark$ & $\times$ & $\times$ & $\times$ & $\times$ & $\times$ & $\times$ & $\times$ \\
  \multicolumn{2}{c}{$E_{1}$}  & 2 & $x,y;xz,yz$ & 1 & 0, $(-1,+1)$ & $\checkmark$ & $\checkmark$ & $\checkmark$ & $\times$ & $\times$ & $\times$ & $\times$ & $\times$ & $\times$ \\
  \multicolumn{2}{c}{$E_{2}$}  & 2 & $x^2-y^2, xy$ & 2 & 0, $(-1,+1)$ & $\checkmark$ & $\checkmark$ & $\checkmark$ & $\checkmark$ & $\checkmark$ & $\times$ & $\times$ & $\checkmark$ & $\checkmark$ \\
\hline\hline
$\mathbf{H}=6$ & $\mathbf{M}=6m'm'$ &  &  &  &  & & & & & &  \\
\hline
$A$ & $DA$ & 1 & $z;x^2+y^2,z^2$ & 6 & 0 & $\times$ & $\checkmark$ & $\times$ & $\checkmark$ & $\times$ & $\checkmark$ & $\checkmark$ & $\times$ &$\times$ \\
$B$ & $DB$ & 1 & $-$ & 3 & 0 & $\times$ & $\checkmark$ & $\times$ & $\times$ & $\times$ & $\times$ & $\times$ & $\times$ & $\times$ \\
$\phantom{,}^{1}E_{1}$& $D^{\phantom{,}1}E_{1}$ & 1 & $x - \text{i}y$, $xz - \text{i}yz$ & 1 & $0$ & $\checkmark$ & $\checkmark$ & $\checkmark$ & $\times$ & $\times$ & $\times$ & $\times$ & $\times$ & $\times$
\\
& & & & & $-1$ & $\checkmark$ & $\checkmark$ & $\checkmark$ & $\times$ & $\times$ & $\times$ & $\times$ & $\times$ & $\times$
\\
$\phantom{,}^{2}E_{1}$ & $D^{\phantom{,}2}E_{1}$ & 1 & $x + \text{i}y$, $xz + \text{i}yz$ & 1 & $0$ & $\checkmark$ & $\checkmark$ & $\checkmark$ & $\times$ & $\times$ & $\times$ & $\times$ & $\times$ & $\times$ \\
& & & & & $+1$ & $\checkmark$ & $\checkmark$ & $\checkmark$ & $\times$ & $\times$ & $\times$ & $\times$ & $\times$ & $\times$ \\
$\phantom{,}^{1}E_{2}$ & $D^{\phantom{,}1}E_{2}$ & 1 & $(x^2-y^2)+2\text{i}xy$ & 2 & $0$ & $\checkmark$ & $\checkmark$ & $\checkmark$ & $\checkmark$ & $\checkmark$ & $\times$ & $\times$ & $\times$ & $\checkmark$ \\
& & & & & $-1$ & $\times$ & $\times$ & $\times$ & $\times$ & $\times$ & $\times$ & $\times$ & $\times$ & $\times$ \\
$\phantom{,}^{2}E_{2}$ & $D^{\phantom{,}2}E_{2}$ & 1 & $(x^2-y^2)-2\text{i}xy$ & 2 & $0$ & $\checkmark$ & $\checkmark$ & $\checkmark$ & $\checkmark$ & $\checkmark$ & $\times$ & $\times$ & $\checkmark$ & $\times$ \\
& & & & & $+1$ & $\times$ & $\times$ & $\times$ & $\times$ & $\times$ & $\times$ & $\times$ & $\times$ & $\times$ \\
\hline
\end{tabular}
\end{ruledtabular}
\end{table*}

In addition to energy and crystal-momentum conservation, pseudo-angular-momentum (PAM) conservation can further constrain the selection rules for first-order phonon Raman scattering. Tatsumi \textit{et al.} derived the corresponding PAM-transfer rule for a phonon mode $\nu$ invariant under an $N_\nu$-fold rotation of the crystal point group \cite{Tatsumi:2018}. Defining the photon helicity change as $\Delta\sigma=\sigma_{\mathrm{out}}-\sigma_{\mathrm{in}}$, the conservation law takes the form
\begin{equation}
    \Delta\sigma
    =
    -m_\nu^{\mathrm{ph}}+N_\nu p,
    \qquad
    p\in\mathbb{Z},
    \label{eq:PAMselection}
\end{equation}
where $\sigma_{\mathrm{in}}$ and $\sigma_{\mathrm{out}}$ are the helicities of the incoming and outgoing photons, respectively, in units of $\hbar$.
Here, $m_\nu^{\mathrm{ph}}$ denotes the PAM quantum number of the phonon mode. The term $N_\nu p$ expresses the conservation of PAM modulo the $N_\nu$-fold rotational symmetry and represents the possible transfer of an integer multiple of $N_\nu\hbar$ to the lattice. The value of $m_\nu^{\mathrm{ph}}$ is determined by the transformation of the phonon displacement eigenvector under the relevant rotation. For a nondegenerate zone-center phonon transforming according to a real irreducible representation, the displacement eigenvector can be chosen real. Such a mode carries no angular momentum and has $m_\nu^{\mathrm{ph}}=0$ within the present convention. This argument does not, however, apply to modes belonging to complex one-dimensional representations.

In the following, we derive the PAM selection rules for the irreducible representations and corepresentations relevant to \mmo{}. In particular, we extend the analysis to complex one-dimensional representations, such as the $D^{1}E_{2}$ and $D^{2}E_{2}$ corepresentations, which were not explicitly considered in Ref.~\cite{Tatsumi:2018}. Complex-conjugate pairs of one-dimensional irreducible representations occur in the crystallographic point groups $4$, $\bar{4}$, $4/m$, $3$, $\bar{3}$, $6$, $\bar{6}$, $6/m$, $23$, and $m\bar{3}$, and in magnetic point groups constructed from them \cite{Bradley:2009}.

To characterize the circular motion associated with a phonon mode, we follow the formulation of Zhang and Niu \cite{Zhang:2014}. For a normalized phonon polarization vector $\boldsymbol{\epsilon}_{\nu,n}$, the component of the PAM along the crystallographic $z$-axis, coinciding with the light propagation direction in our experiment,  carried by mode $\nu$ is
\begin{equation}
\hbar m^{\rm ph}_\nu
=
\hbar
\sum_n
\boldsymbol{\epsilon}_{\nu,n}^{\dagger}
\cdot
\begin{pmatrix}
0 & -\mathrm{i} & 0\\
\mathrm{i} & 0 & 0\\
0 & 0 & 0
\end{pmatrix}
\cdot
\boldsymbol{\epsilon}_{\nu,n},
\label{Eq:PAM}
\end{equation}
where $n$ labels the atoms in the primitive unit cell. A nonzero value of $m^{\rm ph}_\nu$ indicates a chiral phonon whose atomic displacements exhibit a net circular polarization about the $z$ axis. By analyzing the phonon eigenvectors and their contributions from the different Wyckoff positions, one can identify the modes with nonzero angular momentum and assign them to the corresponding irreducible representations. The resulting PAM quantum numbers $m_\nu^{\mathrm{ph}}$ are listed in Table~\ref{tab:selectionrules}.

For \mmo{}, our analysis shows that the finite PAM of the chiral phonon modes originates from the displacements of the atoms occupying the $2a$ and $2b$ Wyckoff positions, namely four Mn and four O atoms in the primitive unit cell. The site-symmetry groups of these positions contain a threefold rotation axis parallel to $z$, allowing circularly polarized displacement patterns about this axis. 

In the paramagnetic phase, these modes belong to the two-dimensional irreducible representations $E_1$ and $E_2$, whose two components form time-reversal-related pairs with opposite PAM quantum numbers, $m_\nu^{\mathrm{ph}}= (-1,+1)$, as listed in Table~\ref{tab:selectionrules}. In the magnetically ordered phase, the $E_1$ and $E_2$ representations decompose into the corepresentations $D^{1}E_1$, $D^{2}E_1$, $D^{1}E_2$, and $D^{2}E_2$. These corepresentations host modes of opposite circular polarization; symmetry does not enforce degeneracy. As follows from Eq.~(\ref{Eq:PAM}), a phonon polarization vector that can be chosen real has vanishing angular momentum, $m_\nu^z=0$ \cite{Zhang:2014,Tatsumi:2018}. In our phonon polarization vectors, the six Mo and twelve O atoms occupying the $6c$ Wyckoff positions do not contribute to the net PAM of these modes. This agrees with the fact that their site-symmetry groups, $m$ in the paramagnetic phase and $m^\prime$ in the magnetically ordered phase, do not contain complex characters.

We make two additional assumptions in applying the PAM selection rules to \mmo{}. First, we restrict the analysis to zone-center phonons within the harmonic approximation. At $\mathbf{k}=0$, the translational part of a screw operation contributes no Bloch phase factor. The screw operation nevertheless permutes symmetry-related atomic sites within the unit cell and rotates their displacement vectors; both effects must be included in its action on the full phonon eigenvector $\boldsymbol{\epsilon}_{\nu}$. Thus, at the zone center, the phonon PAM is determined by the phase acquired by the complete displacement pattern under the screw operation, including both the rotation of the displacement vectors and the permutation of the basis atoms. Because its fractional translation produces no additional phase at the zone center, the sixfold screw axis of \mmo{} is represented by its sixfold rotational counterpart in the point group classification.

Second, we assume that no proper rotation relevant to PAM conservation is combined with an antiunitary operation, as is the case for \mmo{}. These conditions allow us to apply the approach of Tatsumi \textit{et al.} \cite{Tatsumi:2018}. In the magnetic point group $\mathbf{M}=6m^\prime m^\prime$, all antiunitary operations are reflections (i.e., reflections and glide planes in the magnetic space group). Although these antiunitary operations impose additional constraints on the Raman tensors, as summarized in Table~\ref{tab:Ramantensors}, they do not modify the PAM-conservation rule imposed by proper rotations.

The resulting selection rules for the different Raman configurations are summarized in Table~\ref{tab:selectionrules}. They may be represented schematically by combining the conventional Raman-tensor selection rule with the PAM-conservation condition:
\begin{equation}
I_{R,\nu}
\propto
\left|
\mathbf{e}^{*}_{\mathrm{Sc}}
\cdot
\hat{R}_{\nu}
\cdot
\mathbf{e}_{\mathrm{In}}
\right|^{2}
\sum_{p\in\mathbb{Z}}
\delta_{\Delta\sigma+m_{\nu}^{\mathrm{ph}}-N_{\nu}p,\,0},
\label{eq:crosssec}
\end{equation}
where $\mathbf{e}_{\mathrm{In}}$ and $\mathbf{e}_{\mathrm{Sc}}$ are the polarization vectors of the incoming and scattered light, respectively, and $\hat{R}_{\nu}$ is the Raman tensor of phonon mode $\nu$ \cite{Hayes:2004}. The first factor gives the conventional polarization selection rule, whereas the Kronecker delta enforces PAM conservation modulo the $N_{\nu}$-fold rotational symmetry. The symmetry-adapted Raman tensors for the crystallographic point group $6mm$ and the magnetic point group $6m^\prime m^\prime$ are given in Table~\ref{tab:Ramantensors}. They determine the symmetry-allowed Raman channels associated with the basis functions listed in Table~\ref{tab:selectionrules}. Combining these tensor constraints with PAM conservation yields the selection rules listed in the final six columns of Table~\ref{tab:selectionrules} for the different Raman configurations expressed in Porto notation.

One can associate Eq.~(\ref{eq:crosssec}) with Eq.~(2) in the work by Tatsumi and coworkers \cite{Tatsumi:2018}. In a microscopic treatment, the effective Raman tensor $\hat{R}$ arises from electron-photon coupling, whereas PAM conservation applies to the complete scattering process including the electron-phonon interaction \cite{Tatsumi:2018}.

The main question is whether PAM conservation imposes additional selection rules beyond those obtained from the crystallographic Raman tensors. In an ideal collinear backscattering geometry, the circular components of the incoming and scattered photons correspond to a transferred angular momentum of $0$ or $\pm2\hbar$ along the scattering axis. Based on Eq.~(\ref{eq:crosssec}) and Table~\ref{tab:selectionrules}, the only crystallographically Raman-active modes in \mmo{} for which this constraint may produce an additional exclusion are the paramagnetic $E_2$ and ordered $D^1E_2$ and $D^2E_2$ modes with $m_\nu^{\mathrm{ph}}=\pm1$.

In the paramagnetic phase, the two components of each $E_2$ doublet are related by time reversal, leading to degeneracy. Consequently, one may take their real linear combinations with vanishing angular momentum $m_\nu^{\mathrm{ph}}=0$, which can satisfy the PAM selection rule for a helicity change of $\pm2$ through a rotational Umklapp process with $N_\nu=2$, so theory allows the observation of the paramagnetic $E_2$ modes.

Within the magnetic point group description of the ordered phase, $E_2$ doublets decompose as $ E_2 \rightarrow D^1E_2+D^2E_2$, leading to non-degenerate phonon modes that may carry finite angular momentum. One may therefore expect changes in the Raman spectra in general, concerning the positions and the splitting of the paramagnetic peaks, and the selection rules by PAM conservation may become observable. However, these effects may be too weak to be resolvable in experiments. The absence of morphic effect suggests that the lattice vibrations and magnetic degrees of freedom are decoupled. This is a necessary condition for the applicability of spin group description and may support the altermagnetic features as $\mathbf{M} \rightarrow 0$. As we present in the next sections, we observe neither a resolvable splitting nor additional selection rules upon entering the
magnetically ordered phase in \mmo{}.

\begin{table*}[t]
\caption{\label{tab:xy_xx}
Experimental Raman excitation frequencies in \mmo{} (in cm$^{-1}$) measured in  $z(yx)\bar{z}$ and  $z(xx)\bar{z}$ configuration, as well as $z(\sigma_+\sigma_-)\bar{z}$ and $z(\sigma_-\sigma_+)\bar{z}$ configuration in the paramagnetic phase at 85~K and in the magnetically ordered phase at 5~K. Mode assignment is made by comparison with calculated phonon eigenfrequencies obtained from \textit{ab initio} calculations for the 13 expected $E_2(i)$ modes $i=1,\dots,13$ and to the nine expected $A_1(j)$ modes $j=1,\dots,9$ (see Tab.~\ref{tab:calculatedfreq} for all calculated eigenfrequencies). Assignment of the mode at 292~cm$^{-1}$ in the last row as a longitudinal optical (LO) $A_1$ mode is tentative as discussed in the text.}
\begin{ruledtabular}
\begin{tabular}{cccccc|cc|c}

\multicolumn{6}{c|}{this work} &
\multicolumn{2}{c|}{Ref.~\cite{Stanislavchuk:2020}} &

\multicolumn{1}{c}{Mode}   \\

\multicolumn{2}{c}{$z(xx)\bar{z}$} &
\multicolumn{2}{c}{$z(xy)\bar{z}$} &
\multicolumn{2}{c|}{$z(\sigma_+\sigma_-)\bar{z}$, $z(\sigma_-\sigma_+)\bar{z}$} &
\multicolumn{2}{c|}{$z(xy)\bar{z}$}&
\multicolumn{1}{c}{assignment} 
\\
 
85 K & 5 K & 85 K & 5 K & 85 K & 5 K &  85 K & 5 K & $E_{2}$(i)  (calc.) \\

\hline
64.1 & 63.3 & 64.2 & 63.5 & 64.0 & 63.5 & - & - & 58  \\
143  & 143  & 143  & 143  & 143  & 143  & 140 & 140 & 143 \\
185  & 184  & 185  & 184  & 185  & 184  & 184 & 183 & 180 \\
216  & 215  & 216  & 215  & 215  & 215  & 214 & 214 & 210 \\
266  & 266  & 266  & 266  & 266  & 266  & 265 & 265 & 261 \\
325  & 326  & 325  & 326  & 325  & 326  & 323 & 324 & 318 \\
345  & 345  & 345  & 345  & 345  & 345  & 343 & 343 & 342 \\
-    & -    & -    & -    & -    & -    & 442 & 442 & 424 \\
-    & -    & -    & -    & -    & -    & 461 & 461 & 449 \\
-    & -    & -    & -    & -    & -    & 474 & 474 & 465 \\
-    & -    & -    & -    & -    & -    & 514 & 514 & 507 \\
-    & -    & -    & -    & -    & -    & 555 & 555 & 552 \\
-    & -    & -    & -    & -    & -    & 733 & 733 & 726 \\
\hline
204  & 204  & -    & -    & -    & -    & -   & -   & $A_1(1)$ (calc.) at 203  \\
\hline \hline
292  & 292  & -    & -    & -    & -    & -   & -   & $A_1(2)$ (LO) \\ 
\end{tabular}
\end{ruledtabular}
\end{table*}

\section{Experimental results and discussion}

\subsection{Raman scattering of optical phonons}

\begin{figure}[htb]
    \centering
    \includegraphics[width=\linewidth]{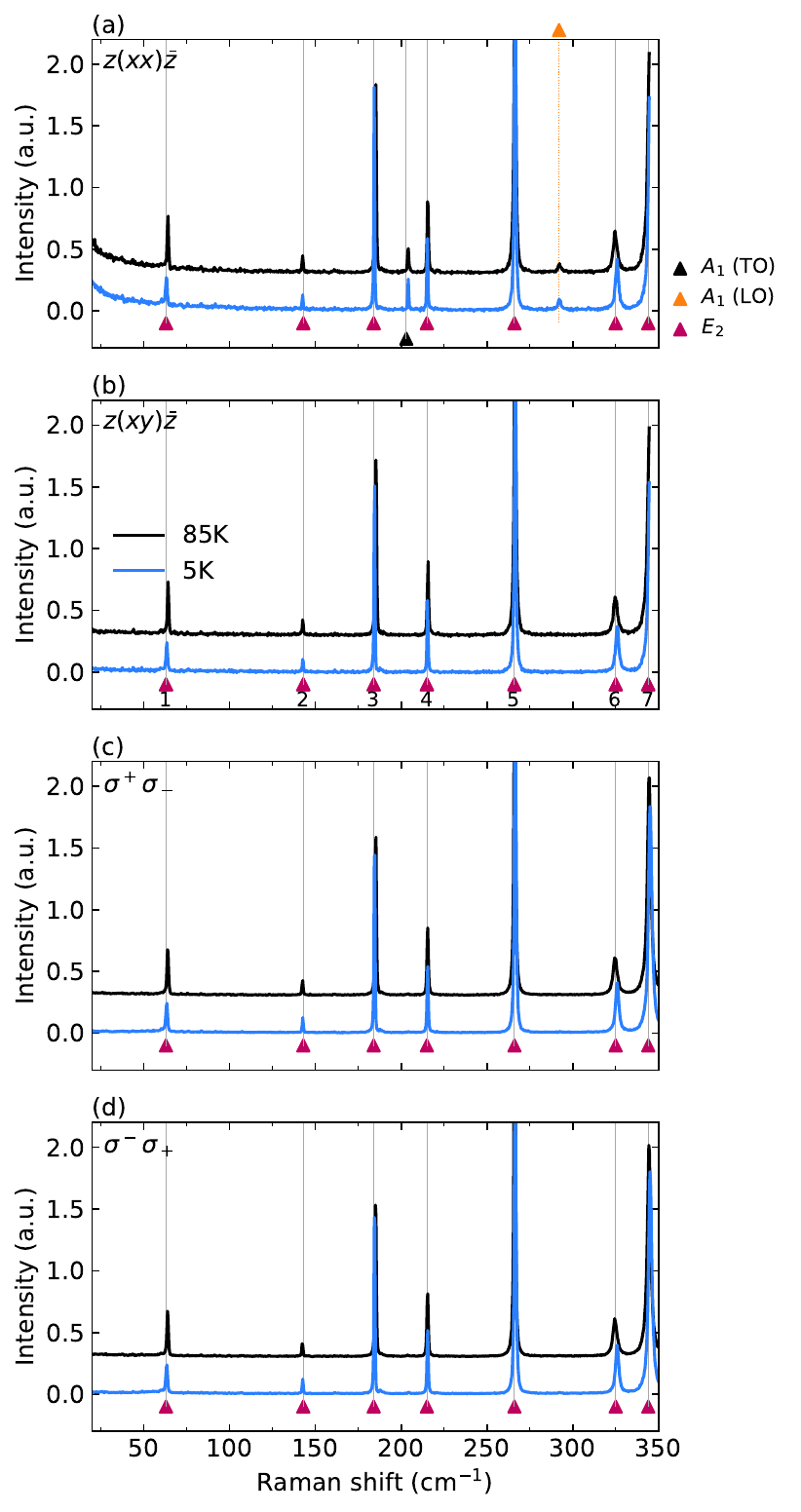}
    \caption{Comparison of Raman spectra at 85 K (black) and 5 K (blue) in  scattering configurations (a) $z(xx)\bar{z}$ allowing for the observation of $E_2$ and $A_1$ modes, (b) $z(xy)\bar{z}$ corresponding to the irreducible $E_2$ representations of the paramagnetic phase. 
    The determined eigenfrequencies of phonon modes expected for each configuration in the paramagnetic phase are indicated by upward triangles at the bottom of the individual panels. Symbols at the top of the panels indicate modes different from expected modes of the paramagnetic phase as described in the text. (c) and (d): Comparison of Raman spectra in crossed circular polarization configurations $z(\sigma_+\sigma_-)\bar{z}$ and $z(\sigma_-\sigma_+)\bar{z}$ at 5 K and 85~K.}
    \label{fig:Raman}
\end{figure}

Following the discussion of possible morphic effects and the correspondig selection rule given by Eqs.~\ref{Eq:irreps} and \ref{Eq:coreps}, we reinvestigated the temperature dependence of the Raman active $E_2$ modes across the magnetic ordering transition  in the Raman channels $z(xx)\bar{z}$, $z(xy)\bar{z}$, $z(\sigma_+\sigma_-)\bar{z}$ and $z(\sigma_-\sigma_+)\bar{z}$. The corresponding spectra are shown in Fig.~\ref{fig:Raman} and the eigenfrequencies obtained by fitting the spectra with Lorentzian lineshapes are compared to the calculated eigenfrequencies in Tab.~\ref{tab:xy_xx}. As we focused in our experiments on the low-frequency regime in order to observe the so-far evasive lowest-lying $E_2$ modes, we restricted the range to below 350~cm$^{-1}$. However, we additionally list the eigenfrequencies of all Raman modes reported previously for the configuration $z(xy)\bar{z}$ by Stanislavchuk \textit{et al}.~\cite{Stanislavchuk:2020} to compare all observable eigenfrequencies with our calculations. A comparison of our calculations to the observed IR- and Raman active modes by Stanislavchuk in the Raman configuration channels $y(zz)\bar{y}$  and $y(xz)\bar{y}$ is provided in App.~\ref{app:alleigenfreqs}. An improved agreement between the experimental and calculated frequencies compared to Ref.~\cite{Stanislavchuk:2020} is due to the use of DFT+$U$ in phonon calculations.

In Fig.~\ref{fig:Raman}(a) and (b) the Raman spectra for the configuration $z(xx)\bar{z}$ and $z(xy)\bar{z}$, respectively, are shown at 85~K in the paramagnetic state and 5~K in the magnetically ordered state. The temperatures were chosen to match the ones used in the previous Raman study by Stanislavchuk \textit{et al}.~\cite{Stanislavchuk:2020}. All expected $E_2$ modes in this frequency range are observed in both configurations, including the lowest-lying $E_2(1)$ with an eigenfrequency of about 63.5~cm$^{-1}$ at 5~K. Almost no shift in eigenfrequencies was observed between the two temperatures, indicating a weak coupling between the spin and lattice degrees of freedom in \mmo. In $z(xx)\bar{z}$ configuration the $A_1$ modes are allowed in addition to the $E_2$ modes. Their calculated eigenfrequencies are given in Tab.~\ref{tab:calculatedfreq}. Only the lowest-lying  $A_1(1)$ mode was observed in our experiment. The overall agreement of the observed and calculated eigenfrequencies is very good and all expected phonon modes could be identified. It is noteworthy to mention that the calculated eigenfrequencies for the $A_1$ and $E_1$ in Tab.~\ref{tab:calculatedfreq} are in good agreement with previously reported experimental values obtained by Raman scattering and IR reflectivity, too \cite{Stanislavchuk:2020}.

Moreover, an additional mode at 292~cm$^{-1}$ is observed in the $z(xx)\bar{z}$ channel at both temperatures, which does not correspond to any of the calculated phonon eigenfrequencies. We attribute the origin of this mode tentatively to the longitudinal optical eigenfrequency of mode $A_1(2)$, based on the fact that several longitudinal $A_1$ modes have been identified in the isostructural materials \cmo{} and \fmo{}, too \cite{Schilberth:2026}. The occurrence of longitudinal modes in these materials has been discussed in terms of intrinsic resonant Raman scattering effects \cite{Schilberth:2026}. Resonant Raman effects can always occur, if the laser frequency is larger than the band gap of the material (see Martin and Falicov in Ref.~\cite{Cardona:1983}). This is also the case for \mmo{}, where our transmission measurements in the MIR-/NIR frequency regime provide an estimate of about 0.84~eV (see Fig.~\ref{fig:MMO_absorption} in App.~\ref{app:opticalgap}), which is clearly lower than the used laser wavelength of 633~nm.
Additional Raman modes reported by Stanislavchuk \textit{et al}. \cite{Stanislavchuk:2020} can also be tentatively assigned to such longitudinal and transverse $A_1$ modes (see Tab.~\ref{tab:zz} and \ref{tab:xz}), confirming a similar pattern as in \cmo{} and \fmo{}.

In Fig.~\ref{fig:Raman}(c) and (d) we show Raman spectra taken at 85~K and 5K for the crossed circular-polarization channels $z(\sigma_+\sigma_-)\bar{z}$ and $z(\sigma_-\sigma_+)\bar{z}$ in order to compare with the derived selection rules in Tab.~\ref{tab:selectionrules}. The mode $A_1(1)$ is allowed in $z(xx)\bar{z}$ both by symmetry of the Raman tensor and PAM conservation, for $z(\sigma_+\sigma_-)\bar{z}$ and $z(\sigma_-\sigma_+)\bar{z}$ actually both factors in Eq.~\ref{eq:crosssec} are zero and the $A_1$ and $DA$ modes are forbidden in these configuration. With respect to the predicted selection rules for the $D\phantom{,}^{2}E_{2}$  and  $D\phantom{,}^{1}E_{2}$ corepresentations, we find no evidence of changes upon entering into the magnetically ordered phase - all modes in the investigated frequency range are observed as in the channels $z(xx)\bar{z}$ and $z(xy)\bar{z}$. We could neither resolve the possible splitting nor observe any circular polarization dependence. This suggests that spin-lattice coupling  in \mmo{} is rather weak and the splitting might be of the order of 10$^{-3}$ as predicted by Sokolov using a Heisenberg Hamiltonian, where the spin-lattice coupling is mediated by the modulation of the exchange integral by lattice vibrations in first order perturbation theory \cite{Sokoloff:1972}. As discussed above, the application of a magnetic field along the $c$-axis is compatible with the magnetic point group $6m'm'$, but might increase the splitting of the $D^{\phantom{,}1}E_1$, $D^{\phantom{,}2}E_1$, $D^{\phantom{,}1}E_2$, and $D^{\phantom{,}2}E_2$ irreducible corepresentations as long as it does not change the collinear spin configuration and the magnetic point group, e.g. by the spin-flop transition in fields above 4~T (see Fig.~\ref{fig:Magnetisation}(b)).

\subsection{THz-time domain magnetospectroscopy}
\begin{figure}[t]
    \centering
    \includegraphics[width=\linewidth]{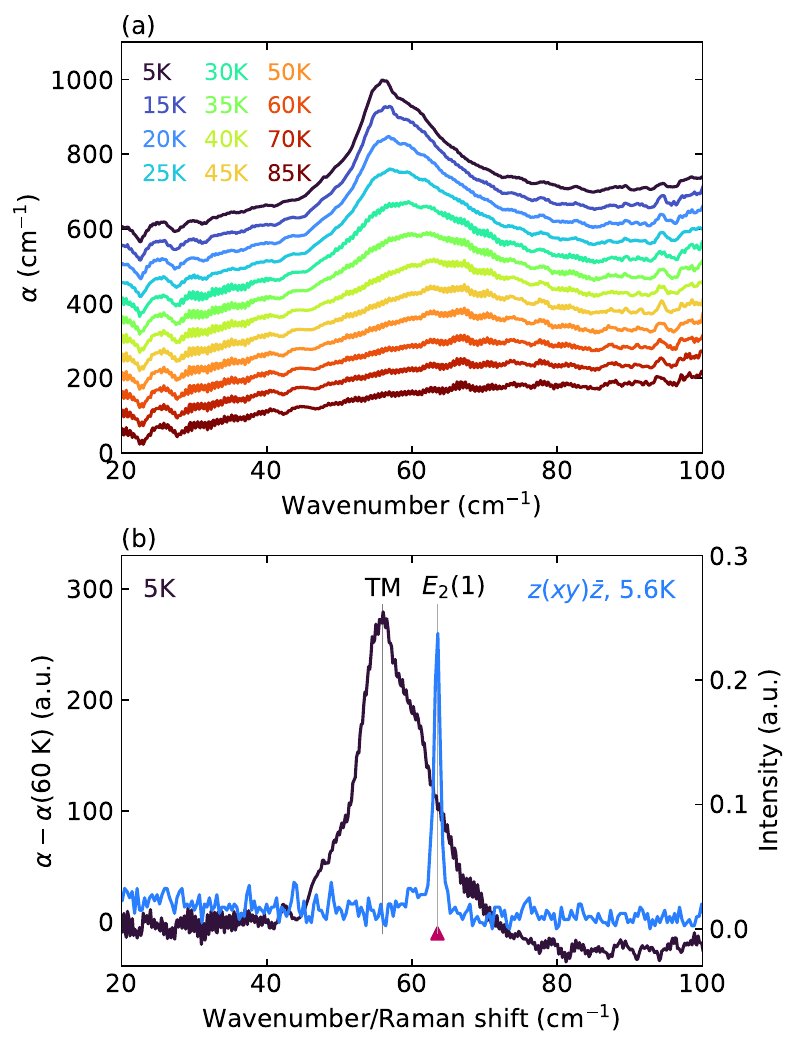}
    \caption{(a) Temperature dependence of the absorption coefficient $\alpha$ in \mmo{} for light polarization configuration $E^\omega,H^\omega\perp c$ crossing into the magnetically ordered state at $T_\text{N}=41$~K. (b) Comparison of the absorption difference $\Delta \alpha=\alpha(5~\text{K})-\alpha(60~\text{K})$ (black) and the lowest-lying $E_2(1)$ Raman mode (blue) in the magnetically ordered state at $T=5$~K.}
    \label{fig:THz_tdep}
\end{figure}
In Fig.~\ref{fig:THz_tdep}(a) we show the temperature dependence of the absorption coefficient of \mmo{} accross the magnetic ordering transition for light polarization configuration $E^\omega,H^\omega\perp c$, where a broad peak called $TM$ emerges at around 60~cm$^{-1}$ from a continuum-like background. The maximum of the $TM$ mode at 4~K is at 58~cm$^{-1}$ (1.74~THz). The peak was previously reported to be only electric-dipole active \cite{Szaller:2020}. As the magnon excitations at the $\Gamma$-point are at lower energies in \mmo{} \cite{Szaller:2020,Liao:2025}, a possible mechanism suggested to account for this broad electric-dipole active THz excitation might be a two-magnon process \cite{Szaller:2020}, where two magnons with opposite propagation vectors can be excited simultaneously by light with $q\approx 0$ \cite{Tanabe:1965,Tanaka:1990,Tanabe:2005,Peedu:2022}. Such processes have first been observed in the rutile antiferromagnets FeF$_2$ \cite{Halley:1965} and MnF$_2$ \cite{Allen:1966} and attributed to an exchange mechanism \cite{Tanabe:1965}, without the need for spin-orbit coupling induced processes as it had been suggested previously \cite{Halley:1965}. Notably, no such broad excitation has been observed in THz spectra of \cmo{} \cite{Reschke:2022,Schilberth:2026} and \fmo{} \cite{Vasin:2024}, where instead phonons are coupled to magnon excitations \cite{Wu:2023,Bao:2023}. In Fig.~\ref{fig:THz_tdep}(b) we compare the broad THz excitation at 5~K after subtraction of the spectrum at 60~K as a non-magnetic background with the lowest-lying Raman mode $E_2(1)$ at 5~K. Although the eigenenergy of the Raman mode matches well with the high-energy flank of the THz excitation, its width is much narrower than the one of the THz excitation and we conclude that the THz excitation \textit{TM} is not strongly influenced by spin-phonon coupling.

\begin{figure*}[htb]
    \centering
    \includegraphics[width=1\linewidth]{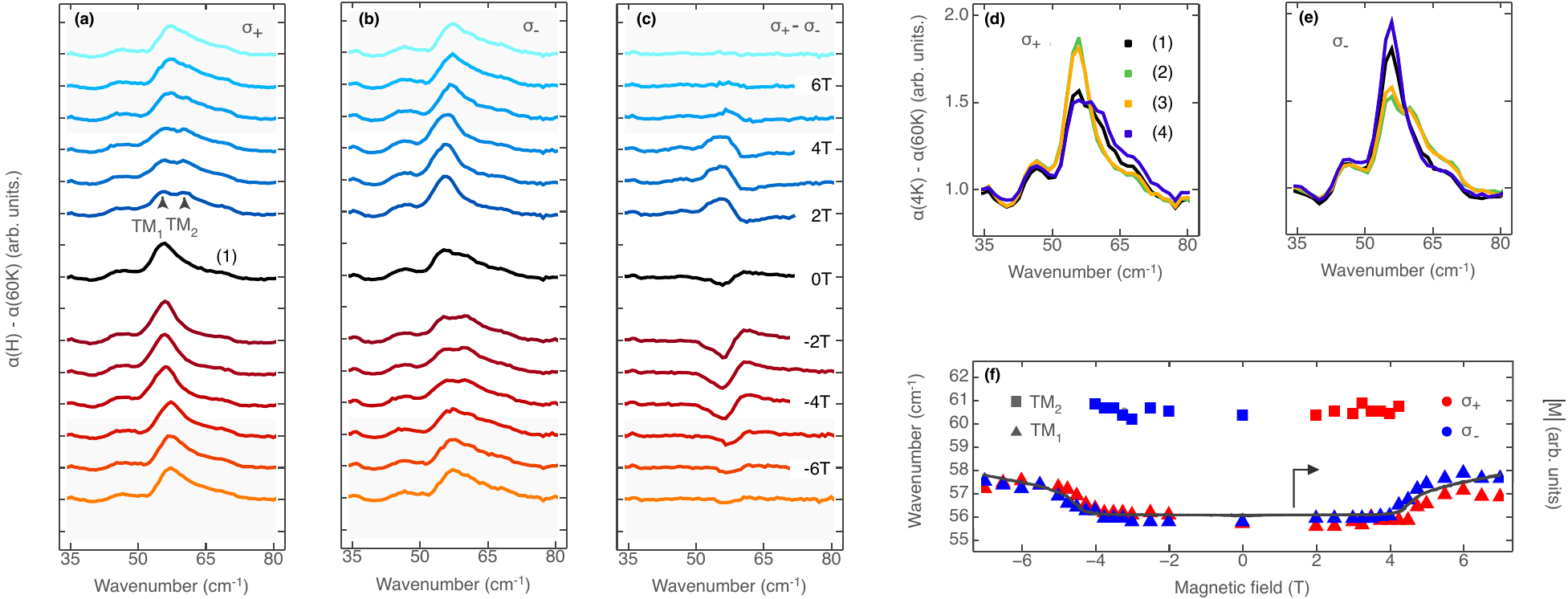}
    \caption{Magnetic field dependence of absorption difference spectra for (a) right circularly polarized light ($\sigma_+$), (b) left circularly polarized light ($\sigma_-$) and (c) the difference of right and left circularly polarized spectra $\sigma_+-\sigma_-$. Panels (d) and (e) show the zero-field absorption spectra for $\sigma_+$ and $\sigma_-$, where (1) corresponds to the first zero-field cooled spectra, (2) to the spectrum taken after the sweep to +7~T, (3) to the one taken after heating to 60~K and zero-field cooling again down to 5~K, and (4) to the one taken after the sweep to -7~T. (f) Magnetic field dependence of the peak frequencies of $TM_1$ and $TM_2$ in comparison to the modulus of the magnetization shown in Fig.~\ref{fig:Magnetisation}(b).}
    \label{fig:THz_Hdep}
\end{figure*}

To further elucidate the origin of this THz excitation, we performed THz-time domain magnetospectroscopy using left and right circularly polarized light in transmission geometry and in fields up to $\pm 7$~T. The sample was first cooled to 5~K in zero field and at each field step from 0 to +7~T, transmission spectra with left and right circularly polarized light were measured. After the magnetic field was ramped down to zero, the sample was heated to 60~K and then cooled down to 5~K and transmission spectra were measured in fields from 0 to -7~T. The corresponding spectra for $\sigma_+$ and $\sigma_-$ circularly polarized light are shown for all fields from -7~T to 7~T in Fig.~\ref{fig:THz_Hdep}(a) and (b), respectively, after subtraction of the spectra at 60~K as a paramagnetic background. Zero-field spectra were taken before and after each field sweep and they are compared in Fig.~\ref{fig:THz_Hdep}(d) and (e). It is clear from this comparison that the zero-field absorption spectra using circularly polarized light depends on its history, which can be expected given the small magnetic hysteresis shown in Fig.~\ref{fig:Magnetisation}(b).  The spectra in zero-field exhibit a clear magnetic circular dichroism, i.e. a difference in the absorption using $\sigma_+$ and $\sigma_-$ circularly polarized light. Either the spectrum is dominated by a single peak $TM_1$ at 57~cm$^{-1}$ or the spectrum shows a two-peak structure consisting of $TM_1$ and the second contribution $TM_2$ with maxima at 56~cm$^{-1}$ and 62~cm$^{-1}$, respectively. Upon application of a magnetic field from 0 to about +4~T, it becomes clear that $TM_1$ appears to be dominant for  $\sigma_-$ circularly polarized light, while the second maximum indicating $TM_2$ seems to be visible only for $\sigma_+$ polarized light. This situation is reversed for negative fields from 0 to -4T. For fields exceeding $\pm$4~T the system enters into the spin-flopped phase and only the low-frequency maximum remains discernible with a high-energy flank that might contain contributions of $TM_2$.

Notably, above the spin-flop transition the spectra exhibit a similar shape suggesting that both $TM_1$  and $TM_2$ are excited independently of the light polarization or field direction and the magnetic circular dichroism is much weaker. In order to highlight field-dependent changes in the spectra, which do not depend on the circular light polarization state, we plot the difference of the spectra taken with right and left circularly polarized light for each field in Fig.~\ref{fig:THz_Hdep}(c). These difference spectra reveal components, which are changing with the helicity of the THz light and should compensate all features unrelated to the polarization state. As expected, the magnetic circular dichroism of the spectra is only related to the frequency range of the $TM_1$ and $TM_2$ contributions and at $\pm 7$~T in the spin-flop phase the flat lines indicate that no MCD is present anymore. In Fig.~\ref{fig:THz_Hdep}(f) we show the peak frequencies of the $TM_1$ and $TM_2$ obtained by parametrizing the two contribution by Lorentzian line shapes. The onset of the spin-flop transition is clearly mirrored by a shift of $TM_1$ to higher frequency, which can be scaled with the modulus of the magnetization shown in Fig.~\ref{fig:Magnetisation}(b). The maximum of $TM_2$ can not be determined unambigously anymore in the spin-flop phase.

\begin{table}[b]
\caption{ \label{tab:neutrons} Magnon energies at the Brillouin zone boundaries taken from Ref.  \cite{Liao:2025}.}
\centering 
\begin{ruledtabular}
\begin{tabular}{lccc}
Q (H,K,L) & Mode & Energy [meV]  & Wavenumber [cm$^{-1}$] \\\hline
$\Gamma_2$ (0,1,0)& $M_1$& 1.9    & 15 \\ 
 & $M_2$& 3.0    & 24 \\ \hline
M (1/2,1,-1)& $M_3$ & 4.5    &35 \\ 
& $M_4$ & 6.2    & 50\\ \hline
K (2/3,2/3,-1)& $M_5$ & 5.4     &44 \\ 
& $M_6$& 6.8    & 55\\ 
\end{tabular}
\end{ruledtabular}
\end{table}

At present, it is unclear how the observed magnetic dichroism and the two-component nature of the THz absorption can be reconciled with the suggested two-magnon process underlying this excitation. We list the Brillouin-zone edge energies of the observed magnon dispersions measured by neutron scattering in Tab.~\ref{tab:neutrons}. While none of these doubled magnon energies corresponds directly to the maxima of $TM_1$ and $TM_2$ branches alone, the combinations of different magnon branches could provide energies in the range of the broad excitation continuum consisting of $TM_1$ and $TM_2$. Whether such two-magnon processes involving different magnon branches are optically allowed at all and could explain the observed magnetic circular dichroism of the absorption, requires, however, a detailed microscopic approach, which is beyond the scope of this paper.

\section{Conclusion}
In summary, we investigated the polar magnet \mmo{} by THz-time domain magnetospectroscopy and Raman scattering and performed a detailed analysis of the pseudo-angular momentum conservation during the Raman process considering the crystallographic and magnetic point groups of the system. Moreover, we calculated the phonon eigenfrequencies for \mmo{} and find them in good agreement with the experimental ones reported both in this work and in a previous study \cite{Stanislavchuk:2020}, which also reported infrared-active modes. As such all expected phonon modes have now been identified. Our Raman measurements do not reveal any activation of previously silent modes or splittings and chiral features of phonon modes upon magnetic ordering, although such processes become allowed by the symmetry lowering due to magnetic order. We conclude therefore, that the optical phonons in \mmo{} follow the same selection rules both in the paramagnetic and magnetically ordered state and do not exhibit morphic effects due to magnetic order. Such a behaviour is expected for magnets which can be described by spin groups with decoupled spatial and spin degrees of freedom \cite{Schilberth:2026}.  In contrast to the sister compounds \fmo{} and \cmo{} the half-filled $d$-shell of the Mn$^{2+}$ ions minimizes spin-orbit coupling effects, which could easily dominate or mask the possible contribution of the secondary altermagnetic multipolar order parameter. On this basis, we conclude that \mmo{} can be an interesting candidate to search for spin-split electronic bands, which are considered the hallmark of altermagnets. Importantly, we observe excitations emerging in the magnetically ordered state. These new excitations, located in the THz spectral range, are only electric-dipole active and show large magnetic circular dichroism below the spin-flop transition. The origin of this behaviour remains unsettled but sets clear restrictions on possible two-magnon processes for this excitation.

\begin{acknowledgments}
 This research was partly funded by the Deutsche Forschungsgemeinschaft (DFG, German Research Foundation)-TRR 360-492547816. THz magneto-spectroscopy by K.P. and A.A.S. were supported by the U.S. Department of Energy (DOE) Grant No. DE-FG02-07ER46382. Experiments at the National Synchrotron Light Source II at Brookhaven National Laboratory were funded by the DOE No. DE-AC9806CH10886. The NSF MPS-ASCEND Award No.~2316535 supported experiments performed by V.A.M. The authors gratefully acknowledge the use of computing resources of the ALCC HPC cluster (Institute of Physics, University of Augsburg).
\end{acknowledgments}

\appendix

\section{Calculated phonon eigenfrequencies for \mmo{} and comparison to Infrared and Raman data}
\label{app:alleigenfreqs}
Here we will discuss all calculated eigenfrequencies for the normal modes as listed in Tab.~\ref{tab:calculatedfreq} with respect to the observed IR- and Raman-active modes as reported by Stanislavchuk \textit{et al.}\cite{Stanislavchuk:2020}. For the nine $A_1$ modes in Tab.~\ref{tab:zz}, a very good agreement between theory and experiment is observed with both the IR-active modes reported for \Epara{} and the Raman-active modes in the Raman configuration $y(zz)\bar{y}$. No signature of an additional activity of the $A_2$ modes as a part of the corepresentation $DA$ in the magnetically ordered phase has been found in the spectra at 5~K. Notably, the lowest lying $A_1(1)$ mode was only observed in the IR channel and not in the $y(zz)\bar{y}$ Raman channel, while in our spectra in $z(xx)\bar{z}$ configuration the mode has been observed (see Fig.~\ref{fig:Raman}(a) and Tab.~\ref{tab:xy_xx}). Instead, mode $A_1(4)$ was only detected in the Raman channel.

\begin{table}[t]
\caption{ \label{tab:calculatedfreq} Calculated phonon eigenfrequencies (in cm$^{-1}$) of the low-temperature magnetic phase ($T=1.7$~K) in \mmo{}.  The calculations do not reveal a  splitting of the $E_1$ and $E_2$ representations in the magnetically ordered state.}
\centering 
\begin{ruledtabular}
\begin{tabular}{cc|cc|cc}
$A_1/DA$ & $A_2/DA$  &$B_1/DB$ & $B_2/DB$    & $E_1/D^iE_1$ & $E_2/D^iE_2$  \\\hline
203 & 141    & 150 & 142   & 157 & 58 \\
256 & 374    & 207 & 406   & 186 & 143\\
371 & 410    & 248 & 380   & 216 & 180\\
439 &     &  365   &       & 270 & 210 \\
449 &     &  438   &       & 301 & 261 \\
549 &     &  477   &       & 337 & 318 \\
631 &     &  565   &       & 425 & 342 \\
709 &     &  633   &       & 447 & 424\\
775 &     &  716   &       & 464 & 448 \\
    &        & 810    &       & 505 & 465 \\
    &        &     &       & 557 & 507 \\
    &        &     &       & 721 & 552 \\
        &        &     &       &  & 726 \\
        \hline
\end{tabular}
\end{ruledtabular}
\end{table}

 In addition, there is a mode observed only in the Raman channel at 668~cm$^{-1}$ at both temperatures and a mode at 844~cm$^{-1}$ appearing in both channels at both temperatures. We tentatively assign the former as the longitudinal optical eigenfrequency of $A_1(7)$ in comparison to a similar observation in  \fmo{} \cite{Schilberth:2026}, where this mode exhibits similar behavior with almost exactly the same transverse and longitudinal eigenfrequencies. The possible origin of the appearance of longitudinal modes in the Raman spectrum is discussed in detail \cite{Schilberth:2026}. The origin of the mode at
 844~cm$^{-1}$ remains unclear, but a mode with similar eigenfrequency has been observed in \fmo{} \cite{Reschke:2020,Stanislavchuk:2020} suggesting an intrinsic process.

The twelve expected $E_1$ modes for light polarization $\mathbf{E}^\omega\parallel a$ and the Raman configuration $y(xz)\bar{y}$ have been identified in a similar fashion by comparison of our calculated values to reported IR- and Raman modes and are listed in Tab.~\ref{tab:xz}. The agreement is equally good as for the other channels, but mode $E_1(12)$ at 729~cm$^{-1}$ was only observed in the IR channel. The additional mode which appears only in the Raman channel is tentatively assigned to the longitudinal optical eigenfrequency belonging to $A_1(9)$, following the same arguments as above for the Raman configuration $y(zz)\bar{y}$.

\begin{table}[t]
\caption{ \label{tab:zz}
Experimental excitation frequencies (in cm$^{-1}$) in \mmo{} of IR-active modes for light polarization \Epara{} and Raman-active  modes in the Raman configuration $y(zz)\bar{y}$ measured at 85\,K and 5\,K as reported in \cite{Stanislavchuk:2020} in comparison to the eigenfrequencies for the nine $A_1$ modes calculated in this work.}
\begin{ruledtabular}
\begin{tabular}{cc|cc|c}
\multicolumn{2}{c|}{IR}          & \multicolumn{2}{c|}{Raman}   &  Mode assignment   \\
 \multicolumn{2}{c|}{\Epara{}}      & \multicolumn{2}{c|}{$y(zz)\bar{y}$}              &  $A_{1}(i)$  \\
85\,K   &  5\,K   & 85\,K & 5\,K    &  $i=1,\dots,9$  (calc.) \\
\hline
204   &  204  & \textit{-}& \textit{-}  & 203  \\
\textit{261}  & \textit{262}  & \textit{252}& \textit{252} & 256 \\
\textit{371}  & \textit{370}  & \textit{369} & \textit{369} & 371\\
\textit{-}  & \textit{-}  & \textit{442} & \textit{442} & 439 \\
\textit{456}  & \textit{456}  & \textit{454} & \textit{454} & 449 \\
\textit{546}  & \textit{547}  & \textit{544} & \textit{544} & 549 \\
\textit{639}  & \textit{640}  & \textit{637} & \textit{637} & 630 \\
\textit{717}  & \textit{718}  & \textit{717} & \textit{717} & 708\\
\textit{780}  & \textit{781}  & \textit{781} & \textit{781} & 775 \\ 
\hline
-  & -  & \textit{668} & \textit{668} & $A_{1}$(7)(LO) \\
\textit{844}  & \textit{842}  & \textit{841} & \textit{841} & not identified \\ 
 \hline
\end{tabular}
\end{ruledtabular}
\end{table}

\begin{table}[h]
\caption{\label{tab:xz}
Experimental excitation frequencies (in cm$^{-1}$) in \mmo{} of IR-active modes for light polarization $\mathbf{E}^\omega\parallel a$ and Raman-active  modes in the Raman configuration $y(xz)\bar{y}$ measured at 85\,K and 5\,K as reported in \cite{Stanislavchuk:2020} in comparison to the eigenfrequencies for the twelve $E_1$ modes calculated in this work.}
\begin{ruledtabular}
\begin{tabular}{cc|cc|c}
\multicolumn{2}{c|}{IR}         & \multicolumn{2}{c|}{Raman}   & Mode   \\
\multicolumn{2}{c|}{$E^{\omega}\parallel a$}   & \multicolumn{2}{c|}{$y(xz)\bar{y}$}  &  assignment    \\
85\,K & 5\,K                  & 85\,K & 5\,K               & $E_{1}$(i) $i=1,\dots,12$ (calc.)\\
\hline
161   & 161     & \textit{159} &   \textit{159} &   157 \\
\textit{188} & \textit{188}   & \textit{187} &   \textit{186} &   186 \\
\textit{221} & \textit{221}   & \textit{219} &   \textit{219} &   216 \\
\textit{274} & \textit{274}   & \textit{272} &   \textit{272} &   270 \\
\textit{307} & \textit{308}   & \textit{306} &   \textit{307} &   301 \\
\textit{340} & \textit{341}   & \textit{339} &   \textit{339} &   337 \\
443   & 443     & \textit{442} &   \textit{442} &   425 \\
\textit{460} & \textit{460}   & \textit{459} &   \textit{459} &   447 \\
\textit{475} & \textit{476}   & \textit{475} &   \textit{475} &   464 \\
\textit{514} & \textit{513}   & \textit{513} &   \textit{513} &   505 \\
\textit{562} & \textit{561}   & \textit{560} &   \textit{560} &   557 \\
\textit{729} & \textit{729}   & \textit{-} &   \textit{-} &   721 \\\hline
- & -   & 781 &   781 &   $A_{1}$(9)(TO) \\

\end{tabular}
\end{ruledtabular}
\end{table}


\section{Irreducible corepresentations of the magnetic point group $6m'm'$}

In Tab.~\ref{tab:6mm} we provide the character tables for the point groups $\mathbf{G}=6mm$ and the halving unitary subgroup $\mathbf{H}=6$, which constitutes the unitary subgroup of the magnetic point group $\mathbf{M}=6 m^\prime m^\prime$. Following the procedure as described by Anastassakis and Burstein \cite{Anastassakis:1972}, we reduce the irreducible representations of $\mathbf{G}=6mm$ with respect to the irreducible representations of $\mathbf{H}=6$ and obtain $A_1, A_2 \rightarrow A$, $B_1, B_2 \rightarrow B$, $E_1\rightarrow ^1E_1 + ^2E_1$, and $E_2\rightarrow ^1E_2 + ^2E_2$ as listed in Tab.~\ref{tab:Coreps_mmo}. As all irreducible representations of $\mathbf{H}=6$ are one dimensional, the corepresentations are immediately obtained as given in the last column of Tab.~\ref{tab:Coreps_mmo}.

\begin{table}[h]
\[
\begin{array}{c|rrrrrrrr}
6mm & E & C^+_6 & C^-_6 & C^+_3 & C^-_3 & C_2 & 3\sigma_d & 3\sigma_v \\
\hline
A_1  & 1 & 1 & 1 & 1 & 1 & 1 & 1 & 1 \\
A_2  & 1 & 1 & 1 & 1& 1 & 1 & -1 & -1 \\
B_1  & 1 & -1 & 1 & 1 & 1& -1 & 1 & -1 \\
B_2  & 1 & -1 & 1 &1 & 1 & -1 & -1 & 1 \\
E_1  & 2 & 1 & 1 & -1 & -1 & -2 & 0 & 0 \\
E_2  & 2 & -1 & -1 & -1 & -1 & 2 & 0 & 0 \\ \hline\hline
6 (C_6) & E & C^+_6 & C^-_6 & C^+_3 & C^-_3 & C_2 &  &  \\
\hline
A  & 1 & 1 & 1 & 1 & 1 & 1 &  &  \\
B  & 1 & -1 & -1 & 1 & 1  & -1 & & \\
^1E_1  & 1 & -e^{ i2\pi/3} & -e^{- i2\pi/3} & e^{- i2\pi/3} & e^{ i2\pi/3} & -1 &  &  \\
^2E_1  & 1 & -e^{ -i2\pi/3} & -e^{ i2\pi/3} & e^{ i2\pi/3} & e^{ -i2\pi/3} & -1 &  &  \\
^1E_2  & 1 & e^{ i2\pi/3} & e^{- i2\pi/3} & e^{- i2\pi/3} & e^{ i2\pi/3} & 1 &  &  \\
^2E_2  & 1 & e^{ -i2\pi/3} & e^{ i2\pi/3} & e^{ i2\pi/3} & e^{- i2\pi/3} & 1 &  &  \\
\end{array}
\]
    \caption{Character table of the crystallographic point groups $\mathbf{G}=6mm$ and the halving subgroup $\mathbf{H}=6$.}
    \label{tab:6mm}
\end{table}

\begin{table}[h]

    \centering
    \begin{tabular}{c|c|c|c}
      $\mathbf{g}=\mathbf{G} \oplus\{E+\tau\}$   & $\mathbf{G}$ & $\mathbf{H}$ &  $\mathbf{M}=\mathbf{H}+\tau(\mathbf{G}-\mathbf{H})$ \\
         & $6mm$ & $6$ &  $6 m^\prime m^\prime$ \\\hline
    DA$_1$  &  A$_1$ & A & DA\\
    DA$_2$  &  A$_2$ & A & DA\\
    DB$_1$  &  B$_1$ & B & DB\\
    DB$_2$  &  B$_2$ & B & DB\\
    DE$_1$  &  E$_1$ & $^1$E$_1$,$^2$E$_1$ & D$^1$E$_1$,D$^2$E$_1$\\
    DE$_2$  &  E$_2$ & $^1$E$_2$,$^2$E$_2$ &  D$^1$E$_2$,D$^2$E$_2$\\
    \end{tabular}
   \caption{Reduction of the irreducible representations of the crystallographic point group $\mathbf{G}=6mm$ with respect to the unitary halving subgroup $\mathbf{H}=6$ to determine the corepresentations of the magnetic point group $\mathbf{M}=6  m^\prime m^\prime$.}
    \label{tab:Coreps_mmo}
\end{table}

The reduced Raman tensors for calculating the scattering cross section $I_{R}\propto \left|\mathbf{e}^*_\text{Sc}\cdot\hat{R}\cdot\mathbf{e}^{\phantom{*}}_\text{In}\right|^2$ are given in a linear polarization basis in Tab.~\ref{tab:Ramantensors} for the irreducible representations of $6mm$ \cite{Hayes:2004} and the corepresentations of $6 m^\prime m^\prime$ \cite{Cracknell:1969}. These symmetry adapted tensors have been tabulated to obtain the corresponding symmetry-adapted functions of the irreducible (co)representation given in Tab.~\ref{tab:selectionrules}. Note that $A_2$ and the $DA$ modes can have a finite cross section originating from the antisymmetric part of their Raman tensors. These contributions are commonly assumed to be negligible, because the leading order effects should stem from the symmetric parts \cite{Birman:1974,Cardona:1983}. Hence, $A_2$ modes are usually considered as silent modes and treated as if they were forbidden by symmetry.
\begin{table}
        \centering
        \caption{Raman tensors of the irreducible (co-)representations for $\mathbf{G}=6mm$ and $\mathbf{M}=6m'm'$ in a linear polarization basis \cite{Hayes:2004,Cracknell:1969}. Note that in \cite{Cracknell:1969} the  nomenclature for $D^{1,2}E_1$ and $D^{1,2}E_2$ is interchanged with respect to ours.}
        \begin{tabular}{l  l}
            \textbf{\( A_1: \)}  &  \textbf{\( DA: \)} \\
             $
             \begin{bmatrix}
                a & 0 & 0 \\
                0 & a & 0 \\
                0 & 0 & b
            \end{bmatrix}
            $ & $
            \begin{bmatrix}
                A & iB & 0 \\
                -iB & A & 0 \\
                0 & 0 & I
            \end{bmatrix}
            $\\
            & \\
            \textbf{\( A_2: \)}  &   \\
             $
             \begin{bmatrix}
                0 & c & 0 \\
                -c & 0 & 0 \\
                0 & 0 & 0
            \end{bmatrix}
            $ & \\
            & \\
             $E_1$:  &  \textbf{\( D^1E_1,D^2E_1: \)} \\
             $
             \begin{bmatrix}
                0 & 0 & 0 \\
                0 & 0 & d \\
                0 & e & 0
            \end{bmatrix}, 
            \begin{bmatrix}
                0 & 0 & -d \\
                0 & 0 & 0 \\
                -e & 0 & 0
            \end{bmatrix}
            $ & $
            \begin{bmatrix}
               0 & 0 & iC \\
                0 & 0 & C \\
                iG & G & 0
            \end{bmatrix},
            \begin{bmatrix}
                 0 & 0 & -iF \\
                0 & 0 & F \\
                -iH & H & 0
            \end{bmatrix}
            $\\
            & \\
            $E_2$:  &  \textbf{\( D^1E_2,D^2E_2: \)} \\
             $
             \begin{bmatrix}
                0 & f & 0 \\
                f & 0 & 0 \\
                0 & 0 & 0
            \end{bmatrix}, 
            \begin{bmatrix}
                f & 0 & 0 \\
                0 & -f & 0 \\
                0 & 0 & 0
            \end{bmatrix}
            $ & $
            \begin{bmatrix}
                D & iD & 0 \\
                iD & -D & 0 \\
                0 & 0 & 0
            \end{bmatrix}, 
            \begin{bmatrix}
                E & -iE & 0 \\
                -iE & -E & 0 \\
                0 & 0 & 0
            \end{bmatrix}
            $\\
        \end{tabular}
        \label{tab:Ramantensors}
\end{table}

Using
$\mathbf{e}_\text{Sc,In}=(x_\text{s,i},y_\text{s,i},z_\text{s,i})$, we obtain in the case of $A_1$ 
\begin{eqnarray}
  \mathbf{e}^*_\text{Sc}  \begin{bmatrix}
    a & 0 & 0 \\
    0 & a & 0 \\
    0 & 0 & b
    \end{bmatrix}\mathbf{e}_\text{In}=a(x^*_\text{s}x_\text{i}+y^*_\text{s}y_\text{i}) + bz^*_\text{s}z_\text{i}
\end{eqnarray}
Hence, the $A_1$ modes should be active in the Raman channels $z(xx)\bar{z}$,$z(yy)\bar{z}$, $x(zz)\bar{x}$, $y(zz)\bar{y}$, but the intensities can be very different for $z(xx)\bar{z}$ and $x(zz)\bar{x}$ due to the different matrix elements $a$ and $b$. The same holds for the corepresentation $DA$.

Similarly, the Raman tensor of the irreducible corepresentation $D\phantom{,}^{1}E_{2}$  yields 
\begin{eqnarray}
  \mathbf{e}^*_\text{Sc}  
    \begin{bmatrix}
        D & iD & 0 \\
        iD & -D & 0 \\
        0 & 0 & 0
    \end{bmatrix}\mathbf{e}_\text{In}&=&D[x^*_\text{s}x_\text{i}-y^*_\text{s}y_\text{i}\nonumber\\ &+& i(x^*_\text{s}y_\text{i} + y^*_\text{s}x_\text{i})],
\end{eqnarray}
in agreement with the characteristic function given in Tab.~\ref{tab:selectionrules}.
Similarly, for $D^1E_1$ we find
\begin{eqnarray}
  \mathbf{e}^*_\text{Sc}  
   \begin{bmatrix}
               0 & 0 & iC \\
                0 & 0 & C \\
                iG & G & 0
            \end{bmatrix}\mathbf{e}_\text{In}&=&iCx^*_sz_i +Cy^*_sz_i +iGx_iz^*_s +Gy_iz^*_s\nonumber \\
            &=& Cy^*_sz_i +Gy_iz^*_s + i(Cx^*_sz_i+Gx_iz^*_s),\nonumber \\
            && \quad
\end{eqnarray}
which is equivalent to the function in  Tab.~\ref{tab:selectionrules}, because the scattering cross section $I_{R}\propto \left|\mathbf{e}^*_\text{Sc}\cdot\hat{R}\cdot\mathbf{e}^{\phantom{*}}_\text{In}\right|^2$
is invariant with respect to any additional phase shift.

\section{MIR/NIR absorption spectra} \label{app:opticalgap}

\begin{figure}[t]
    \centering
    \includegraphics[width=\linewidth]{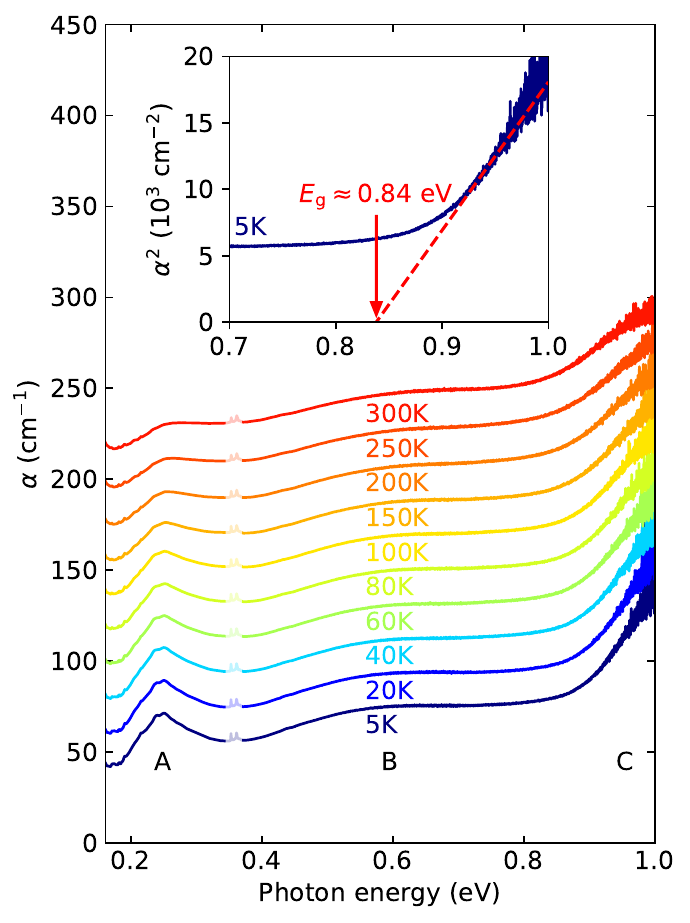}
    \caption{Temperature dependent absorption spectra for light polarization $E^{\omega}\parallel a$ in \mmo{} revealing three excitation bands $A,B,C$ in the MIR/NIR frequency range. In the semi-transparent range two narrow lines appear, which are experimental artifacts and not related to \mmo.  Inset: Plot of $\alpha^2$ vs.~ wave number for the onset of band $C$ at 10~K, yielding an estimate of 1.43~eV for the direct band gap. }
    \label{fig:MMO_absorption}
\end{figure}

The assignment of Raman-active longitudinal optical $A_1$ modes in Tab.~\ref{tab:xy_xx} and Tab.~\ref{tab:zz} was made in analogy to \cmo, where the occurrence was related to resonant-Raman effects, which can occur, if the frequency of the exciting laser is above the band gap of the materials. Therefore, we performed IR transmission experiments in \mmo{} to reveal the optical band gap.
In Fig.~\ref{fig:MMO_absorption}, absorption coefficient spectra  for light polarization $E^{\omega}\parallel a$  in \mmo{} are shown in the mid-infrared (MIR) and near-infrared (NIR) frequency regime for several temperatures crossing the antiferromagnetic ordering transition at $T_\mathrm{N}=40$~K. The absorption coefficient was determined directly from the transmission coefficient $T$ via $\alpha=-(1/d)\ln{T}$, where $d$ is  the  thickness of the sample.

We identify three main broad bands $A,B,C$, which are visible at all temperatures. Upon cooling the intensity of Band $A$ increases and a kind of fine structure becomes visible already below 150~K, far above magnetic ordering. In contrast, band $B$ remains very broad and featureless at all temperatures.
We only observe the onset of the highly absorbing band $C$ before the sample becomes optically opaque for energies larger than 1~eV. We interpret band $C$ as the onset of the semiconducting direct band gap and estimate a gap value of $E_g=0.84$~eV at 5~K from the plot shown as an inset in Fig.~\ref{fig:MMO_absorption}. This value is somewhat lower than the value of 1.43~eV estimated from band structure calculations \cite{Stanislavchuk:2020}.

The origin of bands $A$ and $B$ is not clear at present, but \fmo{} and \cmo{} exhibited similar features at slightly different energy ranges \cite{Park:2021,Vasin:2024,Schilberth:2026}, which were attributed to  multiplet states for the tetrahedral and octahedral sites of  Fe$^{2+}$ and Co$^{2+}$. Hence, we speculate that this holds true also for the Mn$^{2+}$ states in \mmo. Then, the origin of the fine structure of band A may be a signature of vibrationally assisted multiplet states as in many other transition-metal compounds \cite{Deisenhofer:2008,Schmidt:2013,Kocsis:2018a,Park:2021,Vasin:2024,Schilberth:2026}.

\bibliography{Bibliography_MMO}

\end{document}